\documentclass[a4paper,11pt]{article}
\usepackage{pos}
\usepackage[font=footnotesize]{caption}

\title{The ASTRI Mini-Array: in the search for hidden Pevatrons}

\manuallySeparateAuthors
\author*[a]{Martina Cardillo}

\author[b]{ for the ASTRI Project}

\affiliation[a]{INAF -- Istituto di Astrofisica e Planetologia Spaziali,\\
  Via del fosso del Cavaliere 100, 00133 Roma (RM), Italy}

\affiliation[b]{\href{http://www.astri.inaf.it/en/library/}{http://www.astri.inaf.it/en/library/}\\
}

\emailAdd{martina.cardillo@inaf.it}

\abstract{Despite the enormous efforts done in very recent years, both theoretically and experimentally, the basic three questions about the cosmic rays origin remain without clear answers: what are their sources, how are they accelerated, how do they propagate? 
Gamma-ray astronomy plays a fundamental role in this field. Both relativistic protons and electrons can emit in the $\gamma$-ray band through different processes, but only the detection of hadronic $\gamma$-ray emission can probe the acceleration of cosmic rays.
In particular, due to the Klein-Nishina suppression of inverse Compton emission at the highest energies, the detection of $\gamma$-ray emission above 100 TeV was expected to provide firm proof of the acceleration of PeV hadrons. However, the recent results published by the LHAASO collaboration revealed the existence of several PeV sources likely related to PWNe, well known leptonic factories (e.g. the Crab Nebula for all). As a consequence, a $\gamma$-ray detection at PeV energies may no longer be the final proof of hadronic acceleration. However, the limited angular resolution of LHAASO makes associations uncertain and more detailed and deeper studies are needed.
In this context, the ASTRI Mini-Array, with its unprecedented sensitivity and angular resolution at E$>$10 TeV, not only can extend the gamma-ray spectra of candidate Cosmic Ray factories but could help to distinguish emission regions from PWNe and other LHAASO sources, shedding light on the nature of the highest energy emission.}

\FullConference{%
  7th Heidelberg International Symposium on High-Energy Gamma-Ray Astronomy (Gamma2022)\\
  4-8 July 2022\\
  Barcelona, Spain\\}

\def \hess{H.E.S.S.}

\def \gray {$\gamma$-ray}

\def \hess {H.E.S.S.}

\def \astrima {ASTRI Mini-Array}

\begin{document}
\maketitle

\section{The PeVatron Context and the role of ASTRI Mini-Array}
Among the hottest topics in the very high-energy (VHE) astrophysics are cosmic particle acceleration and Cosmic Ray origin. One of the main investigation channels  is non-thermal VHE $\gamma$-ray emission, either produced by electrons (bremsstrahlung or inverse Compton processes) or by protons (pion decay from p-p and p-$\gamma$ interactions). The discrimination between the two main types of processes, leptonic or hadronic, is fundamental to understand particle acceleration phenomena and cosmic rays origin.

In the cosmic rays standard scenario, Galactic sources should be able to accelerate their light component (p and He) at least up to the "knee" energy ($\sim 10^{15}$ eV) (see \cite{Blasi13,Amato14a,Gabici19,Blasi19,Amato2021} for recent reviews). Sources able to accelerate protons up to this energy will be referred to as “hadronic PeVatrons”. Recently, HAWC \cite{Abeysekara20a} and LHAASO \cite{Cao21} have detected several PeVatrons, most of them likely associated to pulsar wind nebulae (PWNe), like the Crab Nebula, and only a few to supernova remnants (SNRs). There is at least one LHAASO detected source, LHAASO J2032+4102, that seems to confirm 100 TeV emission from massive star clusters (MSCs) \cite{Amato2021}. This follows the findings of \hess \citep{HESS16_GC, HESS18_GC}) and then MAGIC \cite{MAGIC20_GC} and HAWC \cite{Abeysekara21} which detected emission around the Galactic Center (GC) and in other MSCs (e.g. Westerlund 1, Cygnus OB2) \cite{Aharonian19} with no evidence of a cut-off up to  $\sim$30 TeV, suggesting that MSCs in the GC may be responsible for the $\gamma$-ray emission. The hadronic or leptonic nature of their emission must be clearly disentangled to assess whether these sources are hadronic PeVatrons. Instead, despite the huge amount of available data, the contribution to cosmic rays acceleration is still largely uncertain.

In this context, the ASTRI Mini-Array will be fundamental thanks to its unprecedented sensitivity and spatial resolution at E > 10 TeV and its wide FoV. The ASTRI Mini-Array is being built at Teide observatory (Tenerife, Canary Islands) thanks to an agreement between INAF and the {\it Instituto de Astrofisica de Canarias, IAC} \cite{Scuderi22}. The ASTRI Mini-Array includes Italian (universities of Perugia, Padova, Catania, Genova and the Milano Polytechnic, INFN sections of Roma Tor Vergata and Perugia,) and international partners (University of S\~ao Paulo with FAPESP in Brasil and the North Western University in South Africa) and it will have the first 3 telescopes operative within 2023 and it will be completed (all the nine telescopes) in 2025.  

The ASTRI Mini-Array is expected to improve the MAGIC and VERITAS sensitivity in the Northern Hemisphere for E > few TeV and to operate for a few years before the full completion of the Cherenkov Telescope Array Observatory (CTAO) North. In particular, its wide FoV ($\sim 10^{\circ}$) with almost homogeneous off-axis acceptance will allow to survey a multi-target field and to enhance serindipitous discoveries. Its angular resolution ($0.05^{\circ}$ at 10 TeV) will be fundamental for a morphological reconstruction of extended sources like SNRs, PWNe and MSCs. Therefore, soon, the ASTRI Mini-Array will have a vast discovery space in the field of extreme $\gamma$-rays, up to 100\,TeV and beyond ~\citep{Vercellone22, Dai22, Saturni22}. 

\section{Candidates Galactic PeVatrons with ASTRI Mini-Array}
In this paper, we present a summary of the ASTRI Mini-Array simulations of expected emission from some selected candidate hadronic PeVatrons. All the details and some cited figures can be found in \citep{Vercellone22}. 

\paragraph{\textbf{Supernova Remnants}}
In the standard paradigm, SNRs are the main contributors of Galactic cosmic rays but the unquestionable evidence of freshly accelerated cosmic rays, hadronic $\gamma$-ray photons with E > 100 TeV, was not found in any SNR. However, Tycho SNR \citep{Park15_Tycho, Archambault17_Tycho} and SNR G106.3+2.7 \citep{Acciari09_Boomerang} have realistic chances to be hadronic PeVatrons. Their $\gamma$-ray spectrum is $\propto E^{-2.3}$ without any clear evidence of a cut-off and, for Tycho, multi-wavelength studies clearly point towards a hadronic origin of this emission \citep{Morlino12_Tycho}. SNR G106.3+2.7 shows two separated emission regions but, although MILAGRO \citep{Abdo07_Boomerang, Abdo09_Boomerang}, HAWC \citep{Albert20_Boomerang}, Tibet AS $\gamma$ \citep{Tibet21_Boomerang}, LHAASO \citep{Cao21} and MAGIC \citep{MAGIC21_Boomerang} detected $\gamma$-rays from the remnant’s region (even up to $\sim$100 TeV) these instruments cannot resolve which one of the two regions is responsible of the highest energy emission because of their low angular resolution.
Only better effective area and sensitivity, as the ones of the ASTRI Mini-Array, can better constrain the spectrum at TeV energies, confirming or disproving the PeVatron nature of these two sources.

\begin{figure}[h!]
    \centering
    \includegraphics[width=0.34\textwidth]{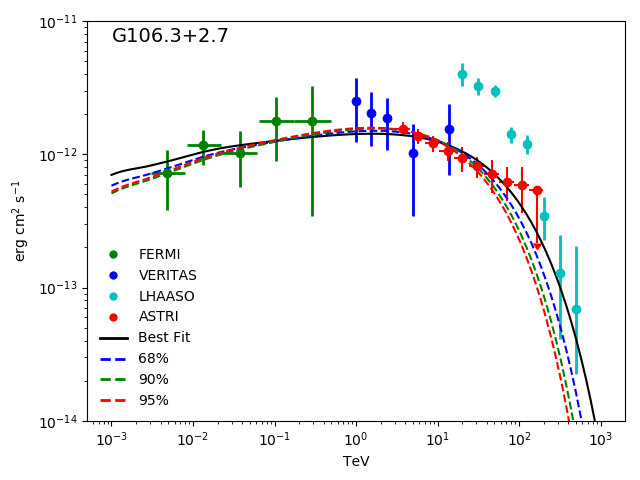}
    \includegraphics[width=0.37\textwidth]{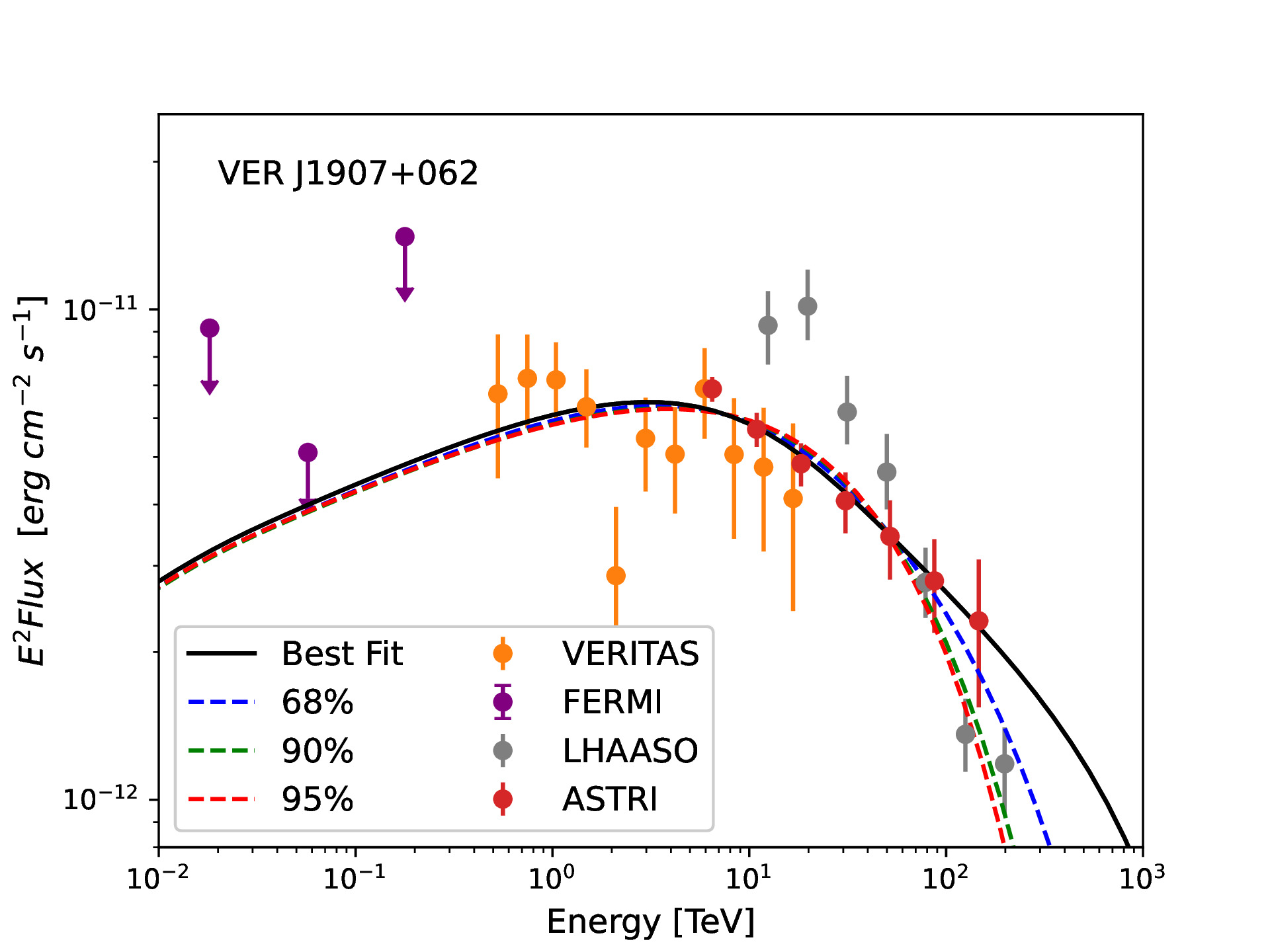}
    \caption{{\it Left panel} - G106.3+2.7: \gray{} data from Fermi-LAT (purple), VERITAS (orange), LHAASO (cyan) and ASTRI Mini-Array (red, 200h). The solid lines show the emission from a proton population with a best-fit cut-off energy of 350 TeV (black) and lower-limit cut-off energy of 250 (blue), 180 (green) and 160 (red) TeV at 68, 90 and 95\% c.l., respectively.{\it Left panel} - VER J1907+062: \gray{} data from Fermi (purple dots), VERITAS (orange dots), LHAASO (grey points) and ASTRI Mini-Array simulations (100h, red dots). The blue, green and red lines show the broken power law fit with a cut-off energy of 1.67, 0.54 and 0.4 PeV, corresponding to 68, 90 and 95\% c.l. respectively.}
      \label{fig:Boomerang+1908}
\end{figure}

In Fig. \ref{fig:Boomerang+1908} left panel and in Fig. 15 of \cite{Vercellone22}, we show simulations of observation of these two hadronic PeVatron candidates with 500 and 200 hours of exposure, respectively. If Tycho should be a PeVatron source, then the ASTRI Mini-Array will be able to detect it (Fig.15 in \cite{Vercellone22}) and also to constrain its cut-off (we can exclude a cut-off below 4 PeV at 68\% c.l.). For Boomerang SNR (Fig. \ref{fig:Boomerang+1908}) a detection with the ASTRI Mini-Array in 200 hr at E$\sim$100 TeV will constrain the proton maximum energy up to E$\sim$500 TeV with very small error bars. But the main strength of the ASTRI Mini-Array will be its angular resolution that will allow us to disentangle different components of the source region at different energies. 

\paragraph{\textbf{Galactic Center (and Massive Star Clusters)}}
The other strong hadronic PeVatron candidate is the GC region (comprises between $\sim 1.5^{\circ}$ in Galactic longitude and $0.2^{\circ}$ in Galactic latitude). In this volume there are the central super-massive black-hole SgrA*, many star-forming regions, pulsars, SNRs, and many other astrophysical potential accelerators that could contribute to the TeV $\gamma$-ray excess detected by H.E.S.S. \citep{HESS16_GC,HESS18_GC}, VERITAS \citep{Archer16_GC} and MAGIC \citep{Ahnen17_GC, MAGIC20_GC}. This emission is perfectly correlated with the gas distribution and shows a hard PL spectrum without evidence of a cut-off at least up to 40 TeV, and lack of variability.

In Figure 17 of \cite{Vercellone22}, it is evident that ASTRI Mini-Array will be able to extend the \hess{} spectrum and, consequently, not only to detect possible particle PeV emission (\gray{} emission above 100 TeV) but also, with the same exposure time as \hess{} (260h), to constraint the cut-off at E < 2 PeV with 95$\%$ c.l.. In addition, the excellent angular resolution of the ASTRI Mini-Array could help to identify the VHE source among several candidates and the very large Field of View (FoV) will allow to map the whole GC region (and other MSC) in a single observation.

\paragraph{\textbf{The Crab Nebula (and other PWNe)}} 
Most of the LHAASO detected PeV sources \citep{Cao21} are likely related to pulsars, well known leptonic factories, and/or their nebulae, PWNe (the Crab Nebula for all), the only galactic sources where we expect to see electrons accelerated up to PeV energies \citep[][]{Amato2021}. Consequently, a \gray{} detection at these energies still cannot be considered the final proof of hadronic acceleration. Furthermore, the limited angular resolution of LHAASO at TeV scale makes associations uncertain and, thus, more detailed and deeper studies are needed. The Crab nebula, being detected also at PeV energies by LHAASO, is an extraordinary and surprising source for the VHE community. The current open question is if its emission is totally due to leptons or if there is also a hadronic component. In order to properly answer this question, the entire SED must be modeled and compared with all available data for different assumptions on the pulsar wind speed and composition \citep{Vercellone22}.

The ASTRI Mini-Array sensitivity will allow us to constrain the hadronic contribution in the Crab Nebula (and similar sources): different hadrons' fraction and energies imply different behavior at the highest energies. In Figure 22 in \cite{Vercellone22} we show that with an exposure of 500h, ASTRI Mini-Array may play a fundamental role to constrain the hadronic contribution to the PeV \gray{} emission of the Crab.

\paragraph{\textbf{Follow-up of LHAASO sources: LHAASO J1908+0621}}
Among other PeVatron candidates, one of the most promising TeV sources is VER J1907+062 (MGRO J1908+06/2HWC J1908+063/LHAASO J1908+0621) \citep[see e.g.][]{Crestan21}. After its discovery by MILAGRO \citep{Abdo07_Boomerang}, this VHE source with a very complex morphology was detected by HAWC up to $\sim$ 100 TeV with no evidence of a cut-off \citep{Abeysekara20a, HAWC22_1908} and recently by LHAASO \citep{Cao21} up to $\sim$ 500 TeV. In spite of a strong correlation between $^{12}$CO emission and the SNR G40.5-0.5, likely connected with VER J1907+062, other possible counterparts cannot be excluded with the current data (see also \citep{Li21_1908}). Also the possible association of the emission with the PWN of PSR J1907+0621 has been considered \citep[e.g.][]{DeSarkar22_1908, Wu22_1908, Chang22_1908}, but in fact there is no consensus on whether the emission is hadronic or leptonic.

The ASTRI Mini-Array, as showed in Figure \ref{fig:Boomerang+1908} right panel, will be able to detect possible PeVatron emission from this source with 200 h of exposure, constraining the cut-off to E<0.96 PeV with 95\% c.l.. An X-ray follow-up of this source and similar ones could be very important in order to constrain the \gray{} emission model and we are moving in this direction through X-ray observations proposals. Moreover, we are working on ASTRI Mini-Array simulations in order to resolve a possible energy-dependent morphology (using a 2-zone models) impossible with the resolution of the current instruments (Crestan et al. in preparation). ASTRI Mini-Array could firmly constrain the origin of the emission from the northern region of VER J1907+062 and thus assess its PeVatron nature.
\section{Conclusions}

The \astrima{} will be a forerunner of the CTAO, with the first three telescopes operative within 2023 and all the nine telescopes will be ready within 2025. Our first simulations show how \astrima{} will contribute in a fundamental way to the cosmic rays study thanks to its unprecedented performance at the TeV and multi-TeV energy bands. We are working on further simulations in order to assess the improvements that are expected thanks to the very good angular resolution of this instrument.

Taking into account all the well known occurrence of unfavorable conditions (bad weather, maintenance operations, Canary "calima"\footnote{This is the name given by local people to the Saharan dust tha covers the islands, affecting air quality and visibility}), we expect to have on average $\sim$1800 h/yrs available for observations, also thanks to the possibility to operate in the presence of the Moon \citep{Lombardi22proc}. The \astrima{} has a great potentiality to answer the question "What are the sources of Galactic cosmic rays?" in the very near future.

\acknowledgments 
This work was conducted in the context of the ASTRI Project. We gratefully acknowledge support from the people, agencies, and organisations listed here: \href{http://www.astri.inaf.it/en/library/}{http://www.astri.inaf.it/en/library/}. This paper went through the internal ASTRI review process. 

{\scriptsize
\bibliographystyle{JHEP}
\bibliography{ID330_Martina_Cardillo.bib}}

\providecommand{\href}[2]{#2}\begingroup\raggedright\begin{thebibliography}{10}

\bibitem{Blasi13}
P.~{Blasi}, \emph{{The origin of galactic cosmic rays}},
  \href{https://doi.org/10.1007/s00159-013-0070-7}{\emph{\aapr} {\bfseries 21}
  (2013) 70} [\href{https://arxiv.org/abs/1311.7346}{{\ttfamily 1311.7346}}].

\bibitem{Amato14a}
E.~{Amato}, \emph{{The origin of galactic cosmic rays}},
  \href{https://doi.org/10.1142/S0218271814300134}{\emph{International Journal
  of Modern Physics D} {\bfseries 23} (2014) 1430013}
  [\href{https://arxiv.org/abs/1406.7714}{{\ttfamily 1406.7714}}].

\bibitem{Gabici19}
S.~{Gabici}, C.~{Evoli}, D.~{Gaggero}, P.~{Lipari}, P.~{Mertsch}, E.~{Orlando}
  et~al., \emph{{The origin of Galactic cosmic rays: Challenges to the standard
  paradigm}},
  \href{https://doi.org/10.1142/S0218271819300222}{\emph{International Journal
  of Modern Physics D} {\bfseries 28} (2019) 1930022}
  [\href{https://arxiv.org/abs/1903.11584}{{\ttfamily 1903.11584}}].

\bibitem{Blasi19}
P.~{Blasi}, \emph{{Acceleration of galactic cosmic rays}},
  \href{https://doi.org/10.1393/ncr/i2019-10166-0}{\emph{Nuovo Cimento Rivista
  Serie} {\bfseries 42} (2019) 549}.

\bibitem{Amato2021}
E.~{Amato} and S.~{Casanova}, \emph{{On particle acceleration and transport in
  plasmas in the Galaxy: theory and observations}},
  \href{https://doi.org/10.1017/S0022377821000064}{\emph{Journal of Plasma
  Physics} {\bfseries 87} (2021) 845870101}.

\bibitem{Abeysekara20a}
{\scshape HAWC Collaboration} collaboration, \emph{Multiple galactic sources
  with emission above 56 tev detected by hawc},
  \href{https://doi.org/10.1103/PhysRevLett.124.021102}{\emph{Phys. Rev. Lett.}
  {\bfseries 124} (2020) 021102}.

\bibitem{Cao21}
Z.~{Cao}, F.A.~{Aharonian}, Q.~{An}, L.X.~{Axikegu}, Bai, Y.X.~{Bai},
  Y.W.~{Bao} et~al., \emph{{Peta–electron volt gamma-ray emission from the
  Crab Nebula}}, \href{https://doi.org/10.1126/science.abg5137}{\emph{Science}
  {\bfseries 373} (2021) 425}.

\bibitem{HESS16_GC}
{HESS Collaboration}, A.~{Abramowski}, F.~{Aharonian}, F.A.~{Benkhali},
  A.G.~{Akhperjanian}, E.O.~{Ang{\"u}ner} et~al., \emph{{Acceleration of
  petaelectronvolt protons in the Galactic Centre}},
  \href{https://doi.org/10.1038/nature17147}{\emph{\nat} {\bfseries 531} (2016)
  476} [\href{https://arxiv.org/abs/1603.07730}{{\ttfamily 1603.07730}}].

\bibitem{HESS18_GC}
{H.~E.~S.~S. Collaboration}, H.~{Abdalla}, A.~{Abramowski}, F.~{Aharonian},
  F.~{Ait Benkhali}, A.G.~{Akhperjanian} et~al., \emph{{Characterising the VHE
  diffuse emission in the central 200 parsecs of our Galaxy with H.E.S.S.}},
  \href{https://doi.org/10.1051/0004-6361/201730824}{\emph{\aap} {\bfseries
  612} (2018) A9} [\href{https://arxiv.org/abs/1706.04535}{{\ttfamily
  1706.04535}}].

\bibitem{MAGIC20_GC}
{MAGIC Collaboration}, V.A.~{Acciari}, S.~{Ansoldi}, L.A.~{Antonelli},
  A.~{Arbet Engels}, D.~{Baack} et~al., \emph{{MAGIC observations of the
  diffuse {\ensuremath{\gamma}}-ray emission in the vicinity of the Galactic
  center}}, \href{https://doi.org/10.1051/0004-6361/201936896}{\emph{\aap}
  {\bfseries 642} (2020) A190}
  [\href{https://arxiv.org/abs/2006.00623}{{\ttfamily 2006.00623}}].

\bibitem{Abeysekara21}
A.U.~{Abeysekara}, A.~{Albert}, R.~{Alfaro}, C.~{Alvarez}, J.R.A.~{Camacho},
  J.C.~{Arteaga-Vel{\'a}zquez} et~al., \emph{{HAWC observations of the
  acceleration of very-high-energy cosmic rays in the Cygnus Cocoon}},
  \href{https://doi.org/10.1038/s41550-021-01318-y}{\emph{Nature Astronomy}
  {\bfseries 5} (2021) 465} [\href{https://arxiv.org/abs/2103.06820}{{\ttfamily
  2103.06820}}].

\bibitem{Aharonian19}
F.~{Aharonian}, R.~{Yang} and E.~{de O{\~n}a Wilhelmi}, \emph{{Massive stars as
  major factories of Galactic cosmic rays}},
  \href{https://doi.org/10.1038/s41550-019-0724-0}{\emph{Nature Astronomy}
  {\bfseries 3} (2019) 561} [\href{https://arxiv.org/abs/1804.02331}{{\ttfamily
  1804.02331}}].

\bibitem{Scuderi22}
S.~{Scuderi}, A.~{Giuliani}, G.~{Pareschi}, G.~{Tosti}, O.~{Catalano},
  E.~{Amato} et~al., \emph{{The ASTRI Mini-Array of Cherenkov telescopes at the
  Observatorio del Teide}},
  \href{https://doi.org/10.1016/j.jheap.2022.05.001}{\emph{Journal of High
  Energy Astrophysics} {\bfseries 35} (2022) 52}.

\bibitem{Vercellone22}
S.~{Vercellone}, C.~{Bigongiari}, A.~{Burtovoi}, M.~{Cardillo}, O.~{Catalano},
  A.~{Franceschini} et~al., \emph{{ASTRI Mini-Array core science at the
  Observatorio del Teide}},
  \href{https://doi.org/10.1016/j.jheap.2022.05.005}{\emph{Journal of High
  Energy Astrophysics} {\bfseries 35} (2022) 1}.

\bibitem{Dai22}
A.~{D'A{\`\i}}, E.~{Amato}, A.~{Burtovoi}, A.A.~{Compagnino}, M.~{Fiori},
  A.~{Giuliani} et~al., \emph{{Galactic observatory science with the ASTRI
  Mini-Array at the Observatorio del Teide}},
  \href{https://doi.org/10.1016/j.jheap.2022.06.006}{\emph{Journal of High
  Energy Astrophysics} {\bfseries 35} (2022) 139}.

\bibitem{Saturni22}
F.G.~{Saturni}, C.H.E.~{Arcaro}, B.~{Balmaverde}, J.~{Becerra Gonz{\'a}lez},
  A.~{Caccianiga}, M.~{Capalbi} et~al., \emph{{Extragalactic observatory
  science with the ASTRI mini-array at the Observatorio del Teide}},
  \href{https://doi.org/10.1016/j.jheap.2022.06.004}{\emph{Journal of High
  Energy Astrophysics} {\bfseries 35} (2022) 91}.

\bibitem{Park15_Tycho}
N.~{Park} and {VERITAS Collaboration}, \emph{{Study of high-energy particle
  acceleration in Tycho with gamma-ray observations}},  in \emph{34th
  International Cosmic Ray Conference (ICRC2015)}, vol.~34 of
  \emph{International Cosmic Ray Conference}, p.~769, July, 2015
  [\href{https://arxiv.org/abs/1508.07068}{{\ttfamily 1508.07068}}].

\bibitem{Archambault17_Tycho}
S.~{Archambault}, A.~{Archer}, W.~{Benbow}, R.~{Bird}, E.~{Bourbeau},
  M.~{Buchovecky} et~al., \emph{{Gamma-Ray Observations of
  Tycho{\textquoteright}s Supernova Remnant with VERITAS and Fermi}},
  \href{https://doi.org/10.3847/1538-4357/836/1/23}{\emph{\apj} {\bfseries 836}
  (2017) 23} [\href{https://arxiv.org/abs/1701.06740}{{\ttfamily 1701.06740}}].

\bibitem{Acciari09_Boomerang}
V.A.~{Acciari}, E.~{Aliu}, T.~{Arlen}, T.~{Aune}, M.~{Bautista}, M.~{Beilicke}
  et~al., \emph{{Detection of Extended VHE Gamma Ray Emission from G106.3+2.7
  with Veritas}},
  \href{https://doi.org/10.1088/0004-637X/703/1/L6}{\emph{\apjl} {\bfseries
  703} (2009) L6} [\href{https://arxiv.org/abs/0911.4695}{{\ttfamily
  0911.4695}}].

\bibitem{Morlino12_Tycho}
G.~{Morlino} and D.~{Caprioli}, \emph{{Strong evidence for hadron acceleration
  in Tycho's supernova remnant}},
  \href{https://doi.org/10.1051/0004-6361/201117855}{\emph{\aap} {\bfseries
  538} (2012) A81} [\href{https://arxiv.org/abs/1105.6342}{{\ttfamily
  1105.6342}}].

\bibitem{Abdo07_Boomerang}
A.A.~{Abdo}, B.~{Allen}, D.~{Berley}, S.~{Casanova}, C.~{Chen}, D.G.~{Coyne}
  et~al., \emph{{TeV Gamma-Ray Sources from a Survey of the Galactic Plane with
  Milagro}}, \href{https://doi.org/10.1086/520717}{\emph{\apjl} {\bfseries 664}
  (2007) L91} [\href{https://arxiv.org/abs/0705.0707}{{\ttfamily 0705.0707}}].

\bibitem{Abdo09_Boomerang}
A.A.~{Abdo}, B.T.~{Allen}, T.~{Aune}, D.~{Berley}, C.~{Chen},
  G.E.~{Christopher} et~al., \emph{{Milagro Observations of Multi-TeV Emission
  from Galactic Sources in the Fermi Bright Source List}},
  \href{https://doi.org/10.1088/0004-637X/700/2/L127}{\emph{\apjl} {\bfseries
  700} (2009) L127} [\href{https://arxiv.org/abs/0904.1018}{{\ttfamily
  0904.1018}}].

\bibitem{Albert20_Boomerang}
A.~{Albert}, R.~{Alfaro}, C.~{Alvarez}, J.R.A.~{Camacho},
  J.C.~{Arteaga-Vel{\'a}zquez}, K.P.~{Arunbabu} et~al., \emph{{HAWC J2227+610
  and Its Association with G106.3+2.7, a New Potential Galactic PeVatron}},
  \href{https://doi.org/10.3847/2041-8213/ab96cc}{\emph{\apjl} {\bfseries 896}
  (2020) L29} [\href{https://arxiv.org/abs/2005.13699}{{\ttfamily
  2005.13699}}].

\bibitem{Tibet21_Boomerang}
{Tibet AS{\ensuremath{\gamma}} Collaboration}, M.~{Amenomori}, Y.W.~{Bao},
  X.J.~{Bi}, D.~{Chen}, T.L.~{Chen} et~al., \emph{{Potential PeVatron supernova
  remnant G106.3+2.7 seen in the highest-energy gamma rays}},
  \href{https://doi.org/10.1038/s41550-020-01294-9}{\emph{Nature Astronomy}
  (2021) }.

\bibitem{MAGIC21_Boomerang}
{MAGIC Collaboration}, V.A.~{Acciari}, S.~{Ansoldi}, L.A.~{Antonelli},
  A.~{Arbet Engels}, D.~{Baack} et~al., \emph{{Resolving the origin of
  very-high-energy gamma-ray emission from the PeVatron candidate SNR
  G106.3+2.7 using MAGIC telescopes}},  in \emph{37th International Cosmic Ray
  Conference (ICRC2021)}, vol.~395 of \emph{International Cosmic Ray
  Conference}, p.~796, Mar., 2022.

\bibitem{Archer16_GC}
A.~{Archer}, W.~{Benbow}, R.~{Bird}, M.~{Buchovecky}, J.H.~{Buckley},
  V.~{Bugaev} et~al., \emph{{TeV Gamma-Ray Observations of the Galactic Center
  Ridge by VERITAS}},
  \href{https://doi.org/10.3847/0004-637X/821/2/129}{\emph{\apj} {\bfseries
  821} (2016) 129} [\href{https://arxiv.org/abs/1602.08522}{{\ttfamily
  1602.08522}}].

\bibitem{Ahnen17_GC}
M.L.~{Ahnen}, S.~{Ansoldi}, L.A.~{Antonelli}, P.~{Antoranz}, C.~{Arcaro},
  A.~{Babic} et~al., \emph{{Observations of Sagittarius A* during the
  pericenter passage of the G2 object with MAGIC}},
  \href{https://doi.org/10.1051/0004-6361/201629355}{\emph{\aap} {\bfseries
  601} (2017) A33} [\href{https://arxiv.org/abs/1611.07095}{{\ttfamily
  1611.07095}}].

\bibitem{Crestan21}
S.~{Crestan}, A.~{Giuliani}, S.~{Mereghetti}, L.~{Sidoli}, F.~{Pintore} and
  N.~{La Palombara}, \emph{{Multiwavelength investigation of the candidate
  Galactic PeVatron MGRO J1908+06}},
  \href{https://doi.org/10.1093/mnras/stab1422}{\emph{Monthly Notices of the
  Royal Astronomical Society} {\bfseries 505} (2021) 2309–2315}.

\bibitem{HAWC22_1908}
A.~{Albert}, R.~{Alfaro}, C.~{Alvarez}, J.R.A.~{Camacho},
  J.C.~{Arteaga-Vel{\'a}zquez}, K.P.~{Arunbabu} et~al., \emph{{HAWC Study of
  the Ultra-high-energy Spectrum of MGRO J1908+06}},
  \href{https://doi.org/10.3847/1538-4357/ac56e5}{\emph{\apj} {\bfseries 928}
  (2022) 116} [\href{https://arxiv.org/abs/2112.00674}{{\ttfamily
  2112.00674}}].

\bibitem{Li21_1908}
J.~{Li}, R.-Y.~{Liu}, D.F.~{de Ona Wilhelmi}, and~{Torres}, Q.-C.~{Liu},
  M.~{Kerr}, R.~{Bühler} et~al., \emph{{Investigating the Nature of MGRO
  J1908+06 with Multiwavelength Observations}},
  \href{https://doi.org/10.3847/2041-8213/abf925}{\emph{\apjl} {\bfseries 913}
  (2021) L33} [\href{https://arxiv.org/abs/2102.05615}{{\ttfamily
  2102.05615}}].

\bibitem{DeSarkar22_1908}
A.~{De Sarkar} and N.~{Gupta}, \emph{{Exploring the Hadronic Origin of LHAASO
  J1908+0621}}, \href{https://doi.org/10.3847/1538-4357/ac6ce5}{\emph{\apj}
  {\bfseries 934} (2022) 118}
  [\href{https://arxiv.org/abs/2205.01923}{{\ttfamily 2205.01923}}].

\bibitem{Wu22_1908}
K.~{Wu}, L.~{Zhou}, Y.~{Gong} and J.~{Fang}, \emph{{Investigating the radiative
  properties of LHAASO J1908+0621}},
  \href{https://doi.org/10.1093/mnras/stac3618}{\emph{\mnras} {\bfseries 519}
  (2022) 1881} [\href{https://arxiv.org/abs/2212.04040}{{\ttfamily
  2212.04040}}].

\bibitem{Chang22_1908}
Z.~{Chang}, X.~{Zhang} and J.-Z.~{Zhou}, \emph{{Pulsars as candidates of LHAASO
  sources J2226+6057, J1908+0621, and J1825-1326}},
  \href{https://doi.org/10.1093/mnras/stac2553}{\emph{\mnras} {\bfseries 516}
  (2022) 4916} [\href{https://arxiv.org/abs/2209.02917}{{\ttfamily
  2209.02917}}].

\bibitem{Lombardi22proc}
S.~{Lombardi}, F.~{Lucarelli} and C.e.a.~{Bigongiari}, \emph{{The data
  processing, simulation, and archive systems of the ASTRI Mini-Array
  project}},  in \emph{\procspie}, vol.~12189 of \emph{Society of Photo-Optical
  Instrumentation Engineers (SPIE) Conference Series}, p.~121890P, Aug., 2022,
  \href{https://doi.org/10.1117/12.2629362}{DOI}.

\end{thebibliography}\endgroup



\providecommand{\href}[2]{#2}\begingroup\raggedright\endgroup

%

\end{document}